%% file: www_paper.tex
\documentstyle[12pt,procsla,proceed,epsfig]{article}

  \include{commands}
\begin{document}
  
  \topskip2.2cm
  \centerline{\large\bf\boldmath KARMEN: PRESENT NEUTRINO-OSCILLATION LIMITS}
  \vspace*{0.3cm}
  \centerline{\large\bf\boldmath AND PERSPECTIVES AFTER THE UPGRADE}
  \topskip0.0cm

  \vspace*{0.6cm}
  \centerline{\footnotesize KLAUS EITEL\footnote{for the KARMEN Collaboration}}
  \baselineskip=13pt
  \centerline{\footnotesize\it Forschungszentrum Karlsruhe,
	Institut f\"ur Kernphysik I} 
  \baselineskip=12pt
  \centerline{\footnotesize\it D-76021 Karlsruhe, Postfach 3640, Germany}
  \centerline{\footnotesize E-mail: klaus@ik1.fzk.de}

  \vspace*{8cm}
  \abstracts{
The neutrino experiment KARMEN is situated at the beam stop neutrino source
ISIS. It provides \numu 's, \nue 's and \numub 's in equal intensities from 
the \pip --\mup --decay at rest. The oscillation channels \numunue\ and 
\numubnueb\ are investigated with a 56\,t liquid scintillation calorimeter at 
a mean distance of $17.6$\,m from the $\nu$--source. 
No evidence for oscillations could be found with KARMEN, resulting in \NCL\ 
exclusion limits of $\sit < 8.5\cdot 10^{-3}$ (\numubnueb) 
and $\sit < 4.0\cdot 10^{-2}$ (\numunue) for $\Dm > 100$\,eV$^2$. 
\\In 1996, the KARMEN neutrino experiment has been upgraded by 
an additional veto system.
Vetoing of cosmic muons passing the 7000\,t massive iron shielding of the 
detector suppresses energetic neutrons from deep inelastic 
scattering of muons as well as from $\mu$--capture in iron. 
Up to 1996, these neutrons penetrating into the detector represented the main 
background for the \numubnueb\ oscillation search.
With an expected reduction of the background rate by a factor of 40 the
experimental sensitivity for \numubnueb\ will be significantly enhanced
towards $\sit \approx 1\cdot 10^{-3}$ for large \Dm .
	}		
   
  \vfill
  \normalsize\baselineskip=15pt
  \setcounter{footnote}{0}

\section{Neutrino Production at ISIS}

\begin{figure}[hbt]
  \centerline{\epsfig{figure=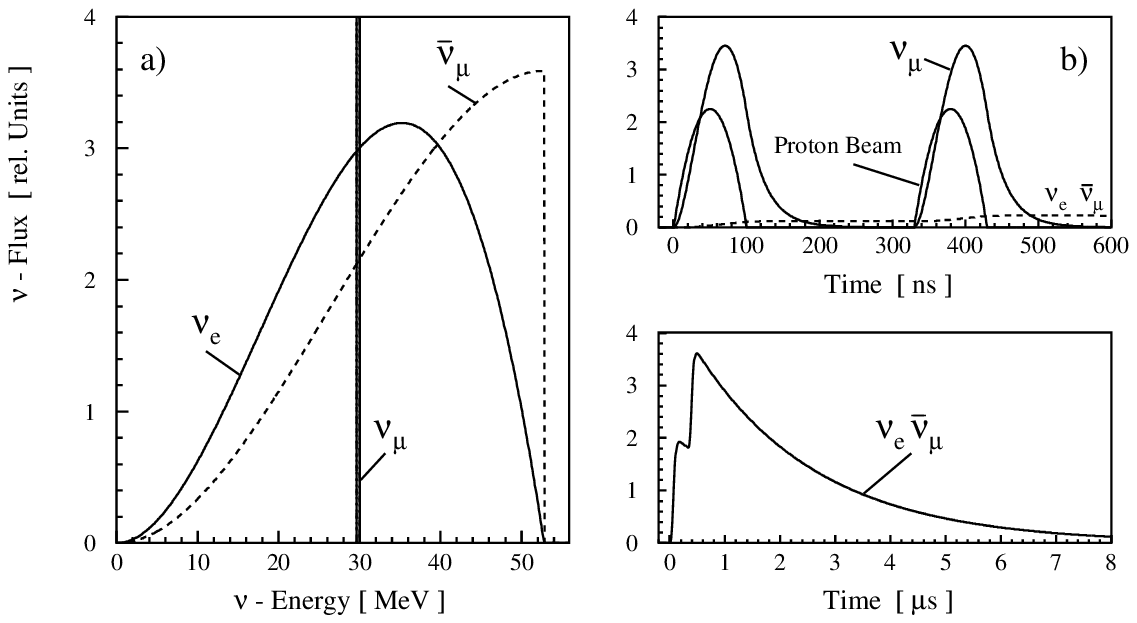,width=15.0cm}}
  \fcaption{Neutrino energy spectra (a) and production times (b) at ISIS.}
  \label{isis_nu}
  \end{figure}

The {\bf K}arlsruhe {\bf R}utherford {\bf M}edium {\bf E}nergy {\bf N}eutrino 
experiment KARMEN
is being performed at the neutron spallation facility ISIS
of the Rutherford Appleton Laboratory. The neutrinos are produced by stopping 
800~MeV protons in a \mbox{beam} dump Ta-$D_{2}O$-target. Neutrinos emerge
from the consecutive decay sequence \pipmup\ and \mupdecay . Thus, ISIS 
represents 
a $\nu$-source with identical intensities for \numu , \nue\ and \numub\ emitted 
isotropically ($\Phi_{\nu}=6.37\cdot 10^{13}$\,$\nu$/s per flavor for 
p-beam current $I_p=200$\,$\mu$A). 
The energy spectra of the $\nu$'s are well defined due to the decay at rest of
both the \pip\ and \mup (fig.~\ref{isis_nu}a).
The \numu 's from \pip --decay are monoenergetic ($E_{\nu}$=30~MeV),
the continous energy distributions of \nue\ and \numub\ up to $52.8$~MeV can be
calculated using the V--A theory. Since \pip\ and \mup\ are stopped in the
small beam dump target, the $\nu$ production region is essentially limited to 
a volume of $\pm 5$\,cm radial to the proton beam and $\pm 10$\,cm along the 
beam axis. With a mean distance source--detector of $L=17.6$\,m
and including the spatial resolution of the detector, the 
uncertainty $\Delta L/L$ for the $\nu$ flight path is less than 1\%.
ISIS therefore ensures that the important experimental parameters $L$ and 
$E_{\nu}$ for $\nu$--oscillations are determined with high precision.

Two parabolic proton pulses of 100\,ns basis width and a gap of 225\,ns are 
produced
with a repetition frequency of 50\,Hz (fig.~\ref{isis_nu}b). The different 
lifetimes of pions 
($\tau$\,=\,26\,ns) and muons ($\tau$\,=\,2.2\,$\mu$s) allow a clear 
separation in time of the \numu -burst from the following \nue 's and 
\numub 's. Furthermore the accelerator duty cycle allows 
effective suppression of any beam-uncorrelated background
by four to five orders of magnitude.

\section{The KARMEN Detector}

The neutrinos are detected in a 56~t liquid scintillation 
calorimeter~\cite{drex1}.
A massive blockhouse of 7000~t of steel in combination with a system 
of two layers of active veto counters provides shielding 
against beam correlated spallation neutron background, suppression
of the hadronic component of cosmic radiation as well as reduction of the flux
of cosmic muons.
The central scintillation calorimeter and the inner veto counters are 
segmented by double acrylic walls with an air gap allowing efficient light 
transport via total internal reflection of the scintillation light at the 
module walls. The event position is determined by the individual module and 
the time difference
of the PM signals at each end of this module. Due to the optimized optical
properties of the organic liquid scintillator and an active volume 
of 96\% for the calorimeter, an excellent energy resolution of 
$\sigma_E=\frac{11.5\%}{\sqrt{E [MeV]}}$ is achieved. In addition, 
Gd$_2$O$_3$ coated 
paper within the module walls provides efficient detection of thermal 
neutrons due to the very high capture cross section of the \Gdng\ reaction.
The KARMEN electronics is synchronized to the ISIS proton pulses to an
accuracy of better than $\pm 2$\,ns, so that the time structure of the 
neutrinos, especially of \numu 's, can be exploited in full detail.
In these proceedings we only present results of the oscillation search and
its results obtained in the first data taking period from 1990--1995,
the investigation of $\nu$--nucleus interactions on \C\ is described 
elsewhere\cite{babod} in detail.
  
\section{Oscillation limits on \numunue\ }

\begin{figure}[hbt]
  \centerline{\epsfig{figure=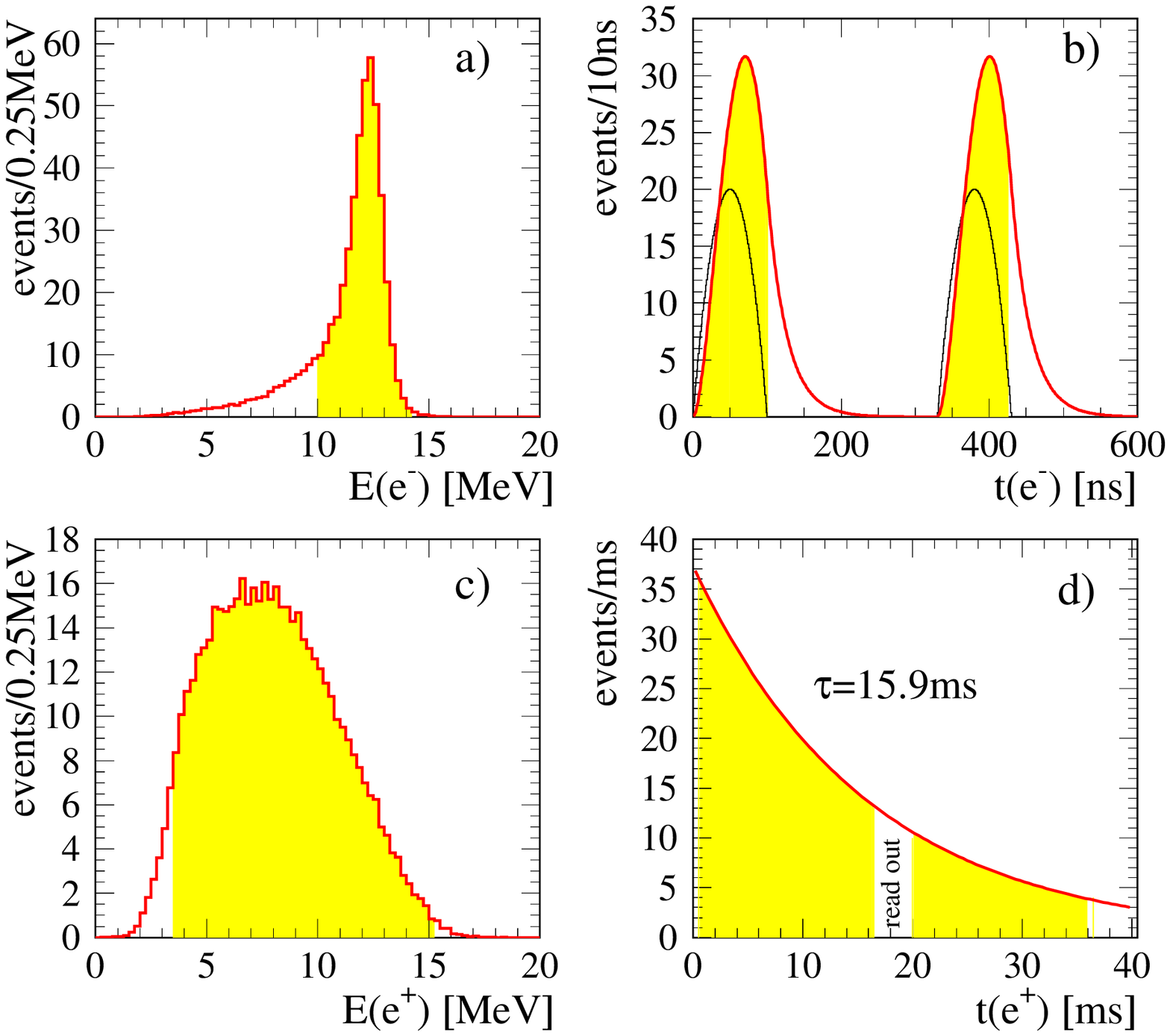,width=11.0cm}}
  \fcaption{Expected signature for \numunue\ full oscillation:\\
	a) simulated MC energy of prompt event; b) proton pulses and time of 
	prompt event relative to ISIS beam-on-target; c) MC energy of 
	sequential event; d) time difference
	between prompt and sequential event; shaded areas show the allowed 
	regions of evaluation cuts.}
  \label{numunue_exp}
  \end{figure}

In the event of \numunue\ oscillations, monoenergetic 30\,MeV \nue 's would 
arise in the \numu\ time window after beam-on-target. 
The \nue\ detection reaction is \excl\ followed by the
$\beta$--decay \Ndecay . One would therefore expect electrons with a peaked
energy spectrum ($E_{e^-}=E_{\nu}-Q=29.8-17.3=12.5$\,MeV, see 
fig.~\ref{numunue_exp}a) within the two \numu\ time pulses 
(fig.~\ref{numunue_exp}b). This characteristic prompt energy signal is
followed by the energy of 
the sequential spatially correlated \ep\ (fig.~\ref{numunue_exp}c) which follows
the \el\ with the typical \N\ decay time (fig.~\ref{numunue_exp}d).
In the later \numu\ time window one can measure the number of \excl\ reactions 
induced by \nue's from \mup -decay in the ISIS target, which can be used
to calculate the expectation of \numunue\ induced CC reactions for 
$\Pmu = 100$\%. The different detection efficiencies and the energy 
dependence of the cross section have to be taken into account to extract the
$\nu$-flux and cross section independent expectation of 187.8 oscillation 
signatures. Applying all cuts
(e.g. $10\le E_{pr}\le 14$\,MeV; $0\le t_{pr}\le 100$\,ns or
$325\le t_{pr}\le 425$\,ns) only 2 sequences remain within the data
taken between July~1992 and December~1995. 
$0.50\pm 0.20$ cosmic induced events contribute to the background
which is dominated by the small contribution of \nue 's from \mup --decay 
within the two 100\,ns long \numu --time intervals after beam-on-target 
($1.76\pm 0.2$). With a total background of $N_{bg}=2.26\pm 0.3$ events,
there is no hint for \numunue\
oscillations and an upper limit of the oscillation probability P of\\
\centerline{$\Pmu < 3.8/187.8=2.0\cdot 10^{-2}$\qquad (\NCL)}
can be extracted. Due to the normalization 
of the full oscillation expectation this result is very reliable. The 
oscillation search is nearly background free so that the sensitivity for 
this oscillation channel is essentially limited by the relatively small 
expectation value for full oscillation, i.e. statistics.

\section{Oscillation limits on \numubnueb\ }

\begin{figure}[hbt]
  \centerline{\epsfig{figure=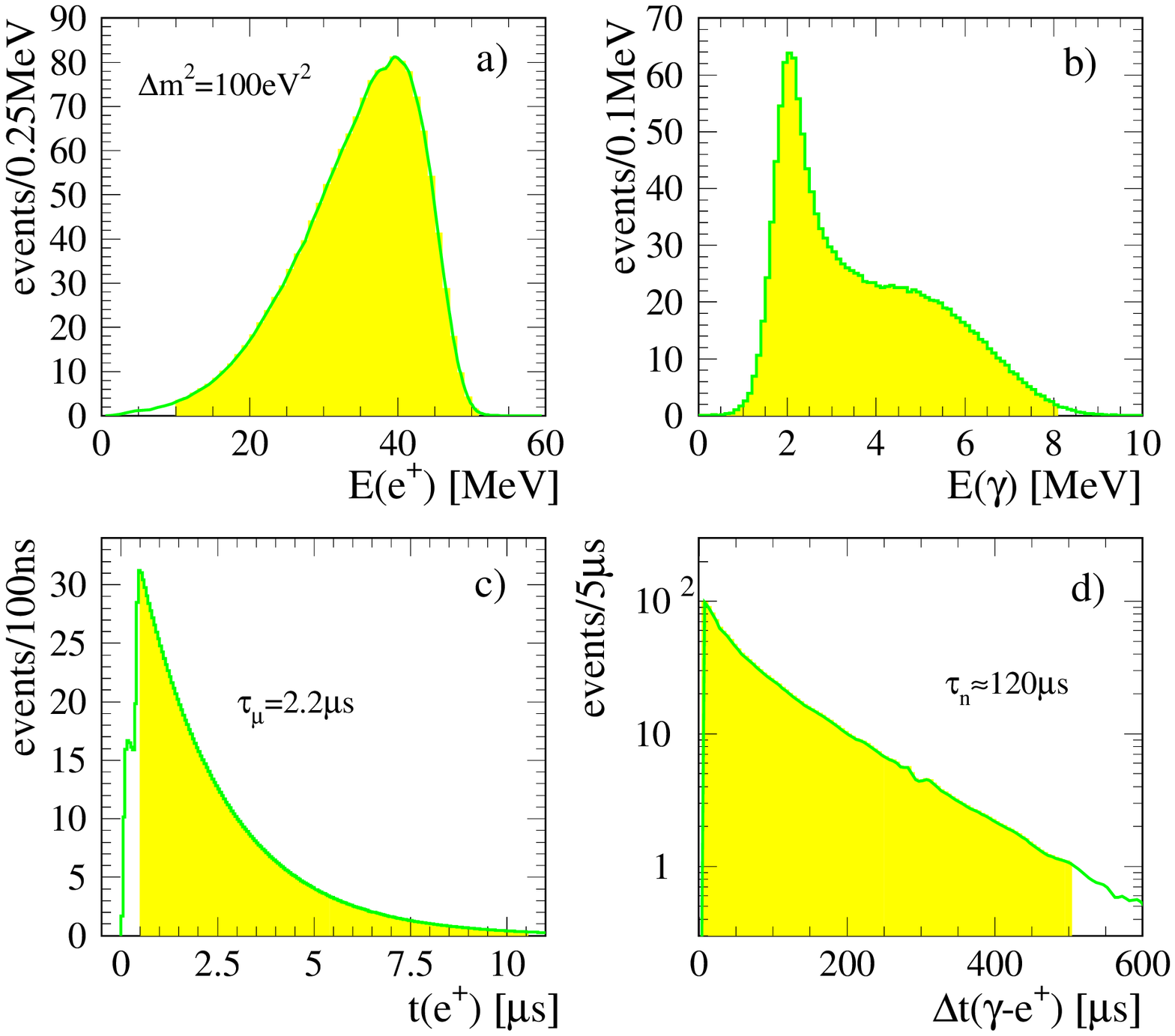,width=11.0cm}}
  \fcaption{Expected signature for \numubnueb\ full oscillation:\\
	a) MC energy of prompt positron for $\Dm = 100$\,eV$^2$; 
	b) energy of sequential $\gamma$'s;
	c) time of prompt event relative to ISIS beam-on-target; 
	d) time difference between prompt $e^+$ and sequential $\gamma$'s; 
	 shaded areas are accepted by evaluation cuts.}
  \label{numubnueb_expect}
  \end{figure}

\begin{figure}[hbt]
  \centerline{\epsfig{figure=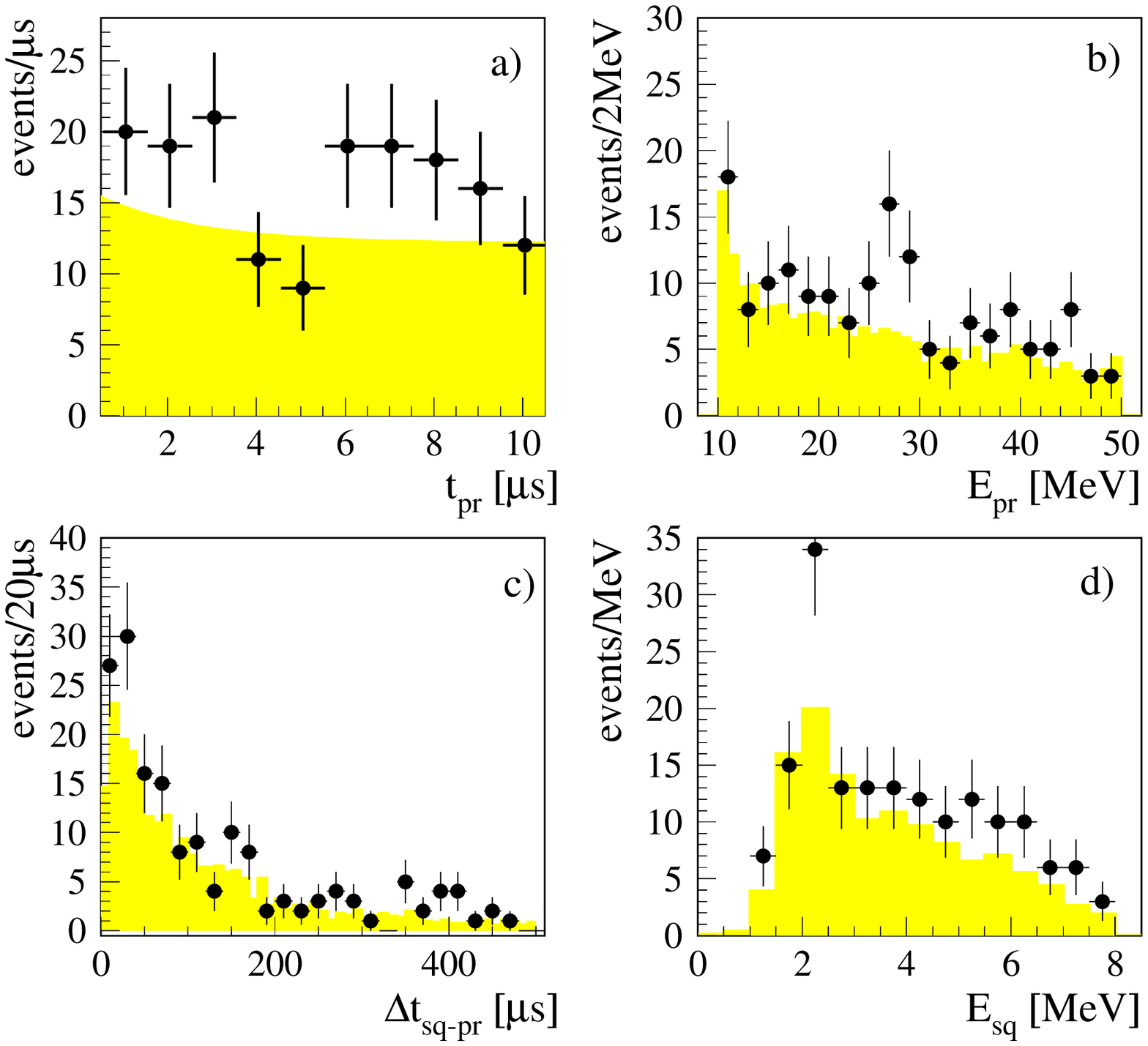,width=11.0cm}}
  \fcaption{Time (a,c) and energy (b,d) distribution of reduced sequences; 
	shaded lines and histograms represent the pre-beam background 
	(12.2 events per $\mu$s) plus \nue -induced CC events and 
	\nueb -contamination.}
  \label{numubnueb_events}
  \end{figure}

The most sensitive mode of the KARMEN experiment for the search of 
$\nu$-oscillations is the \numubnueb\ channel. 
First, \nueb 's are not produced within the ISIS target
apart from a very small contamination of $\nueb /\nue < 6\cdot 10^{-4}$.
The detection of \nueb 's via \CCprot\ would therefore 
indicate oscillations \numubnueb\ in the appearance channel. 
Secondly, the cross section for \CCprot\ with \nueb's from oscillations
($\sigma = 93.6\cdot 10^{-42}$cm$^2$ for $\Dm=100$\,eV$^2$) is about 20 
times larger than for
\excl\ with \nue's from \numunue\ ($\sigma = 4.95\cdot 10^{-42}$cm$^2$), and
the ratio of target nuclei of the scintillator is $H/C=1.767$.
\\The signature for the detection of \nueb 's is a spatially correlated 
delayed coincidence of positrons from \CCprot\ with energies up to 
$E_{e^+}=E_{\nueb}-Q=52.8-1.8=51$\,MeV (fig.~\ref{numubnueb_expect}a) and 
$\gamma$ emission of either of the two neutron capture processes \pnd\ or 
\GdngG\ with $\gamma$ energies of 2.2\,MeV or up to 8\,MeV, respectively
(fig.~\ref{numubnueb_expect}b).
The positrons are expected in a time window of 0.5 to 10.5\,$\mu$s after
beam-on-target (fig.~\ref{numubnueb_expect}c). The neutrons from \CCprot\ are
thermalized and captured typically with $\tau=120$\,$\mu$s 
(fig.~\ref{numubnueb_expect}d). The neutron detection efficiency for the
analyzed data is 28.2\%. The data set remaining after applying all cuts in
energy, time and spatial correlation is shown in fig.~\ref{numubnueb_events}.
A prebeam analysis of cosmic ray induced sequences results in an accumulated
background level of $12.2\pm 0.2$ events per $\mu$s in the prompt 
10\,$\mu$s--window (see fig.~\ref{numubnueb_events}a). 
The actual rate is $16.4\pm 1.3$/$\mu$s which 
corresponds to a beam excess of 2.4\,$\sigma$ compared with the prebeam level
including \nue -induced CC (9 events) and \nueb -contamination (1.7 events).
Although the secondary part of the sequences shows the typical signature of
thermal neutron capture, the prompt time and energy distribution does not 
follow the expectation from \numubnueb\ oscillation with $\Dm = 100$\,eV$^2$.

\begin{figure}[hbt]
  \centerline{\epsfig{figure=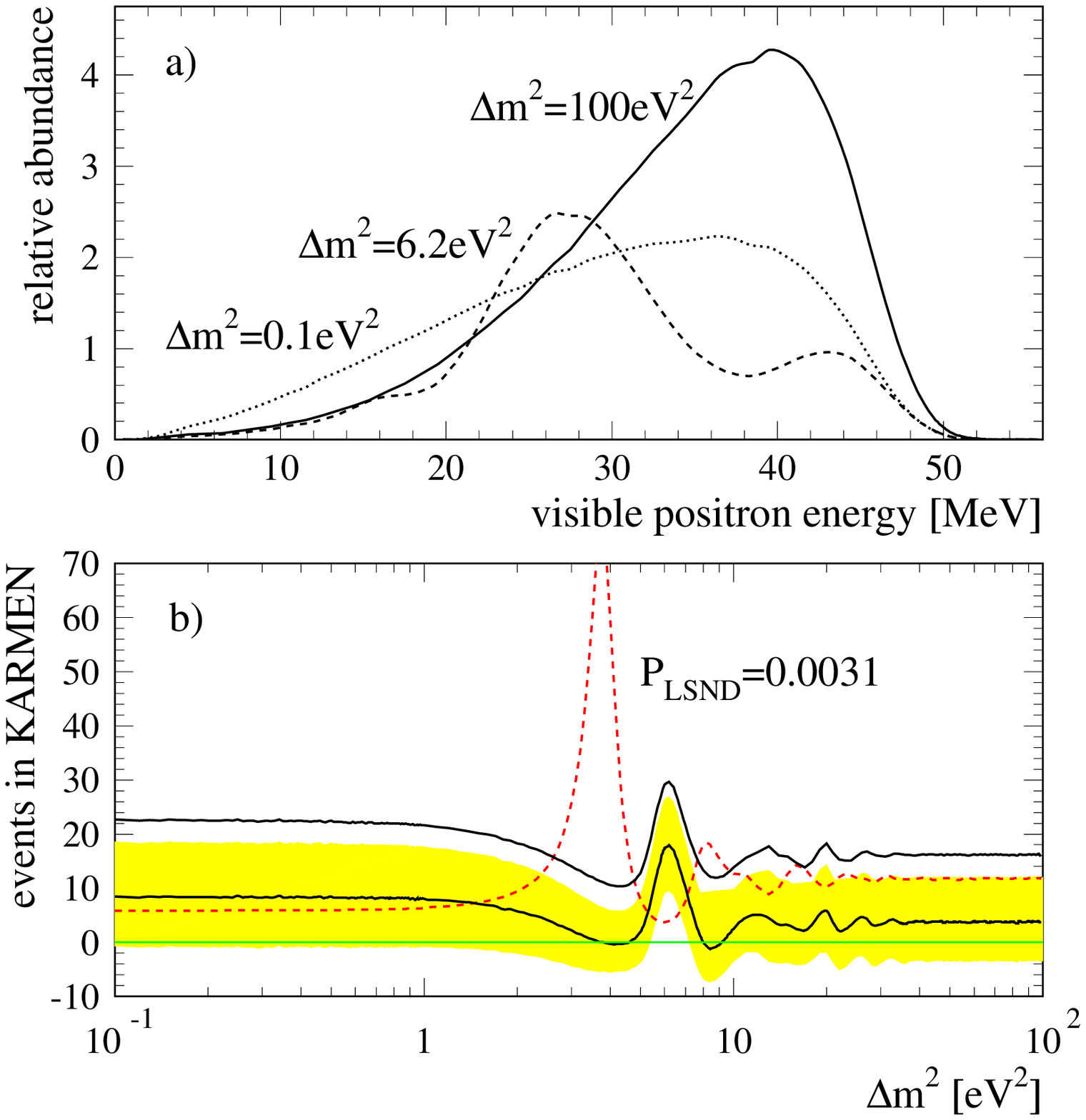,width=11.0cm}}
  \fcaption{a) Examples of expected \ep --spectra (visible energy including
	detector response) for different oscillation parameters \Dm ;
	b) likelihood-fit results depending on \Dm ; the shaded
	area represents the 1$\sigma$--error band around the best fit values
	for $\Dm = 0.01\dots 100$\,eV$^2$; the solid line above the shaded band
	shows the \NCL\ upper limit from KARMEN, the broken line the expected
	event numbers in KARMEN based on the LSND oscillation 
	evidence~\cite{atha}.}
  \label{numubnueb_dm2}
  \end{figure}

To extract a possible small contribution of \numubnueb , the data set is scanned
with a two-dimensional maximum likelihood analysis on time and energy 
distribution of the positrons requiring a 2.2\,$\mu$s exponential time
constant for the \ep\ and a time independent cosmic induced background.
The measurement of the \ep\ energy with spectroscopic quality is highly 
sensitive to changes in the energy spectrum due to the dependence of the
oscillation probability on the mass term \Dm\ (see fig.~\ref{numubnueb_dm2}a). 
The energy distributions of the positrons used in the likelihood analysis 
therefore have been tested with spectra for \Dm\ in the range from 0.01 to 
100\,eV$^2$. 

For most of the investigated parameter range of \Dm\ the likelihood
analysis results in best fit values compatible with a zero signal within
a $1\sigma$ error band (see fig.~\ref{numubnueb_dm2}b). Only for a parameter
region at $\Dm = 6.2$\,eV$^2$ there is a positive signal $2.3\sigma$ above
zero which is not considered as statistically significant. In addition,
this \Dm -value corresponds to the first theoretical oscillation minimum in 
the detector with the lowest possible mean energy of the positrons
(see also fig.~\ref{numubnueb_dm2}a) and represents therefore an extremum
in the likelihood analysis which should be interpreted with special 
precaution.

On this basis of no evidence for oscillations, \NCL\ upper limits for 
oscillation events (fig.~\ref{numubnueb_dm2}b) as well as for the oscillation 
parameters \Dm\ and \sit\ are deduced.
Our result can be compared with an expected signal of 6 up to 76 events for 
\Dm =3.9\,eV$^2$ based on a recently published oscillation 
probability of $\Pmub = 0.0031$ by LSND~\cite{atha}. For large \Dm\ we expect
at KARMEN 1898 detected oscillation events for full oscillation which results
in an upper limit of the mixing angle\\
\centerline{$\sit < 16.3/1898=8.5\cdot 10^{-3}$\quad for large \Dm
	\qquad (\NCL).}

\begin{figure}[hbt]
  \centerline{\epsfig{figure=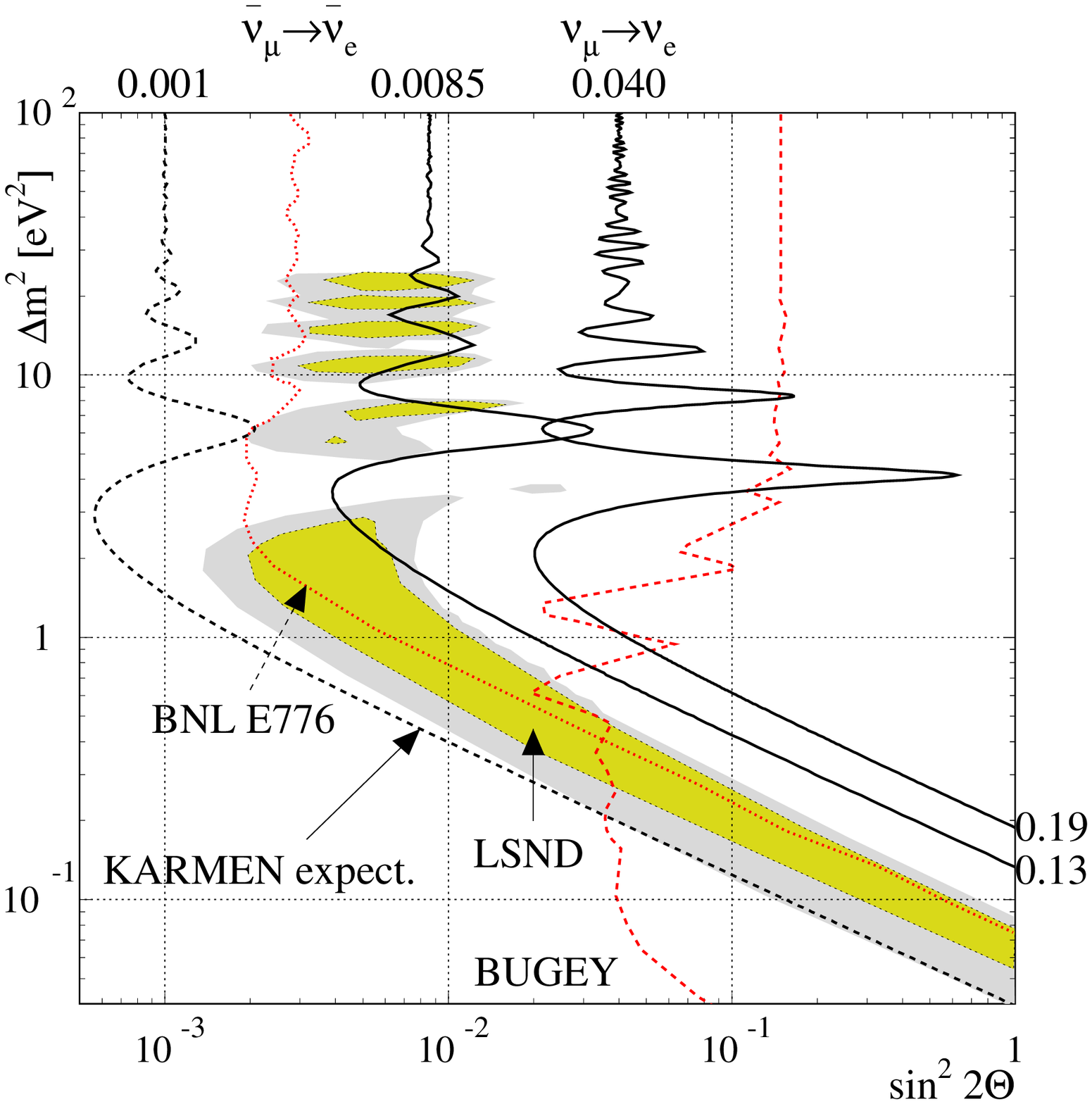,width=12.0cm}}
  \fcaption{\NCL\ exclusion curves and limits for $\Dm = 100$\,eV$^2$, 
	$\sit = 1$ from KARMEN for \numunue\ and \numubnueb\ as 
	well as the expected sensitivity for \numubnueb\ after the upgrade;
	oscillation limits from BNL E776~\cite{bnl} and Bugey~\cite{bug};
	LSND evidence is shown as shaded areas (\NCL\ and \NNCL\ areas 
	respectively).}
  \label{numubnueb_exclu}
  \end{figure}

Fig.~\ref{numubnueb_exclu} shows the KARMEN exclusion curves in the parameter 
space of \Dm\ and \sit\ in a two neutrino flavor oscillation calculation for the
appearance channels \numunue\ and \numubnueb\ in comparison with
other results of $\nu$--oscillation searches at accelerators~\cite{bnl} 
and reactors~\cite{bug}. As the sensitivity for \numubnueb\ of the 
KARMEN experiment is comparable to that of LSND (both experiments expect about
2000 oscillation events for $\Dm > 100$\,eV$^2$ and $\sit = 1$ on their data
sample until 1995), the KARMEN \NCL exclusion curve cannot exclude 
the entire parameter space favoured by the positive result of LSND.

\section{The KARMEN Upgrade}

For at least the next 3 years, no other oscillation experiment will cover the 
whole parameter region favored by LSND. Only the running KARMEN experiment, 
with an improved sensitivity, will be able to crosscheck the evidence 
postulated by LSND. We therefore investigated different scenarios for 
increasing the \numubnueb\ sensitivity.
Whereas the sensitivity for \numunue\ oscillations is essentially limited by
statistics, the KARMEN sensitivity in the \numubnueb\ channel can only be 
substantially increased by the reduction of the small but dominant cosmogenic 
background (see fig.~\ref{numubnueb_events}). 
This background is induced by cosmic muons
stopping or undergoing deep inelastic scattering in the iron shielding which
surrounds the KARMEN detector and veto system. Energetic neutrons emitted in 
these processes can penetrate deep into the detector without triggering the 
veto system, thus producing an event sequence of prompt recoil protons 
followed by the capture of the then thermalized neutrons.

\begin{figure}[hbt]
  \epsfclipon
  \centerline{\epsfig{figure=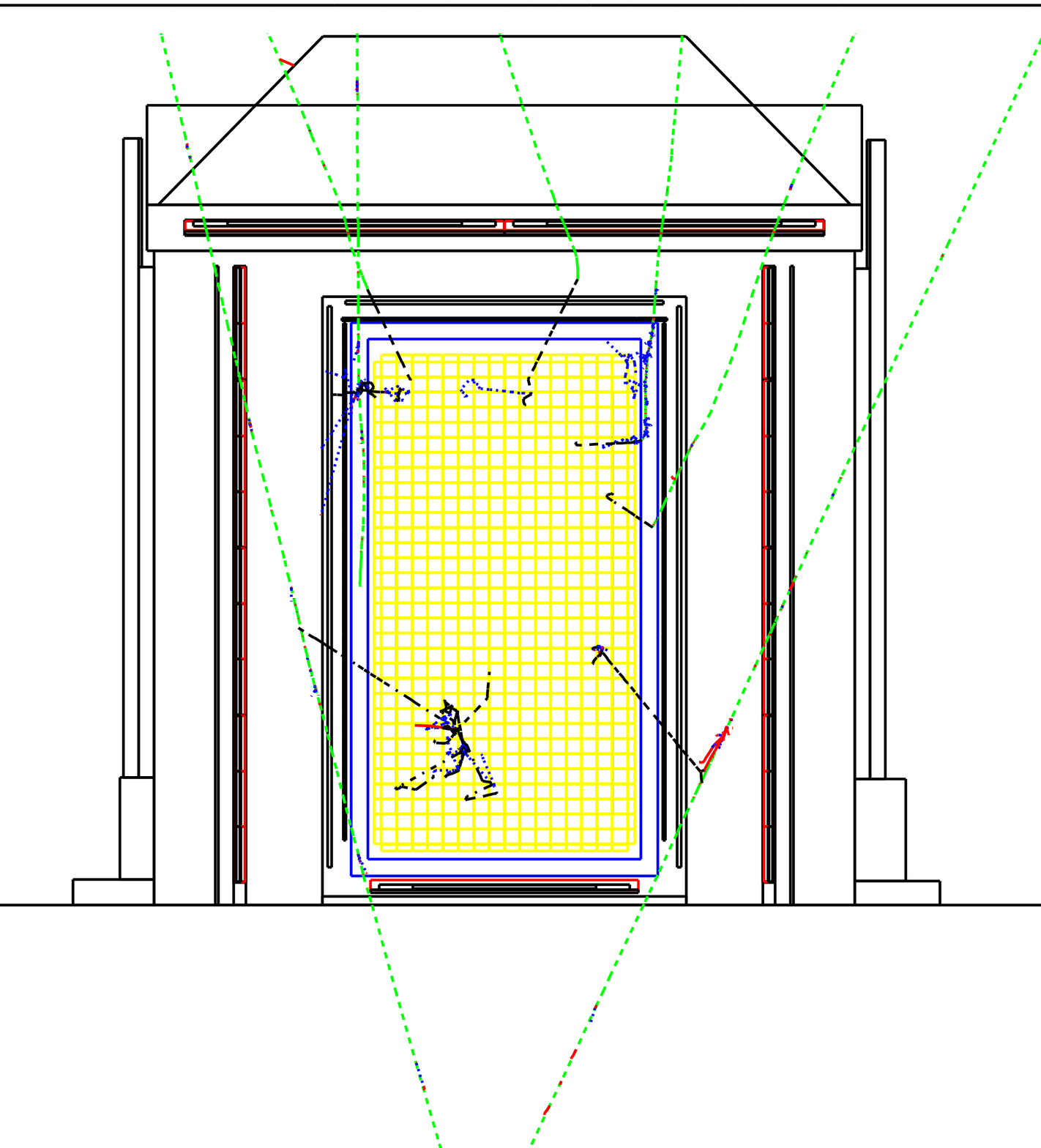,
	width=12.0cm}}
  \fcaption{Cross section of the KARMEN central detector with surrounding 
	shield counters and massive iron blockhouse, now including the 
	additional veto system.
	Cosmic muons passing or stopping in the iron which produce
	energetic neutrons can now be tagged by the new veto scintillators.}
  \label{blockhouse}
  \end{figure}

To tag the original muons in the vicinity of the detector, a further active 
veto layer within the blockhouse has 
been built in 1996, which consists of 136 plastic scintillator bars 
(BICRON BC 412) of lengths up to 4\,m, 65\,cm width and 5\,cm thickness, 
with a total surface of 
300\,m$^2$ covering all sides of the detector (fig.~\ref{blockhouse}).
There is at least 1\,m between the new counter and the existing shield so that
energetic neutrons produced by cosmic muons outside the new veto system have to
travel a path of more than 4 attenuation lengths in iron ($\Lambda=21$\,cm).
This reduces cosmogenic neutrons to a negligible fraction of
less than 1.5\% of the original flux. The new veto system is designed to
reduce cosmogenic sequential background by a factor of at least 40.
This reduction factor is based on detailed background measurements and
extensive GEANT MC simulations of cosmic muons. First preliminary evaluations
of cosmic background in a prebeam window indicate the expected reduction 
factor in the energy region of interest 
when the information of veto hits is included in the analysis.

After two years of measuring time with the new detector configuration, 
the KARMEN sensitivity for \numubnueb\ is expected to exclude the whole 
parameter region of evidence suggested by LSND if no oscillation signal will 
be found (fig.~\ref{numubnueb_exclu}). In that case, mixing angles with
$\sit > 1\cdot 10^{-3}$ will be excluded for large \Dm .
The veto upgrade will also increase the signal to background ratio of single
prong $\nu$-induced events on \C\ and therefore improve the investigation of 
the published anomaly in this time distribution~\cite{anomaly}.

\section{Acknowledgements}

We acknowledge the financial support of the German Bundesministerium f\"ur
Bildung, Wissenschaft, Forschung und Technologie.

\section{References}

  \end{document}

%% file: commands.tex
\newcommand{\C}{\mbox{$^{12}$C}}

\newcommand{\N}{\mbox{$^{12}$N}}

\newcommand{\Ngs}{\mbox{$^{12}$N$_{\rm g.s.}$}}
\newcommand{\B}{\mbox{$^{12}$B}}

\newcommand{\numu}{\mbox{$\nu_{\mu}$}}
\newcommand{\numub}{\mbox{$\bar{\nu}_{\mu}$}}
\newcommand{\nue}{\mbox{$\nu_{e}$}}

\newcommand{\nueb}{\mbox{$\bar{\nu}_{e}$}}

\newcommand{\ep}{\mbox{e$^{+}$}}

\newcommand{\el}{\mbox{e$^{-}$}}
\newcommand{\pos}{\mbox{e$^{+}$}}

\newcommand{\mup}{\mbox{$\mu^{+}$}}

\newcommand{\pip}{\mbox{$\pi^{+}$}}

\newcommand{\mupdecay}{\mbox{\mup\ $\rightarrow\:$ \pos $\!$ + \nue\ + \numub}}

\newcommand{\pipmup}{\mbox{\pip $\rightarrow\:$ \mup + \numu}}

\newcommand{\Ndecay}{\mbox{\Ngs\ $\rightarrow\:$ \C\ + \pos\ + \nue}}

\newcommand{\CCprot}{\mbox{p\,(\,\nueb\,,\,\ep\,)\,n }}
\newcommand{\excl}{\mbox{\C\,(\,\nue\,,\,\el\,)\,\N$_{\rm g.s.}$}}

\newcommand{\numunue}{\mbox{\numu $\rightarrow\,$\nue }}
\newcommand{\numubnueb}{\mbox{\numub $\rightarrow\,$\nueb }}

\newcommand{\Pmu}{\mbox{$P_{\numunue}$ }}
\newcommand{\Pmub}{\mbox{$P_{\numubnueb}$ }}

\newcommand{\NCL}{\mbox{$90\%\,CL$ }}
\newcommand{\NNCL}{\mbox{$99\%\,CL$ }}

\newcommand{\Dm}{\mbox{$\Delta m^2$}}
\newcommand{\sit}{\mbox{$sin^2(2\Theta )$}}

\newcommand{\Gdng}{\mbox{Gd\,(\,n,$\gamma$\,)}}

\newcommand{\GdngG}{\mbox{Gd\,(\,n,$\gamma$\,)\,Gd}}

\newcommand{\pnd}{\mbox{p\,(\,n,$\gamma$\,)\,d}}